\newcommand{\hf}{{_1\over^2}}

\newcommand{\qq}{\begin{equation}}
\newcommand{\qqq}{\end{equation}}
\newcommand{\myeqnarray}[1]{
  \begingroup
  \jot=#1pt
  \arraycolsep=2pt
  \begin{eqnarray}}
\newcommand{\beqnarray}{\myeqnarray{3}}
\newcommand{\eeqnarray}{\end{eqnarray}\endgroup}
\newcommand{\ba}{\begin{array}{cc}}
\newcommand{\ea}{\end{array}}

\documentclass[twocolumn,showpacs]{revtex4}
\usepackage[xdvi]{graphicx}%
\usepackage{dcolumn}
\usepackage{epsfig}
\usepackage{feynmf}

\begin{document}
\draft
\title{Normal heat conductivity in a strongly pinned chain of anharmonic  oscillators.}
\author{Rapha\"el Lefevere${}^{1}$ and Alain Schenkel${}^{2}$}
\affiliation
{${}^1$ LPMA-UFR Math\'ematiques Universit\'e Paris 7, Case 7012,  75251 Paris Cedex 05, France 
\\
${}^2$ Department of Mathematics
P.O. Box 68
FIN-00014 University of Helsinki
Finland}

\date{\today}

\begin{abstract}
We consider a chain of coupled and strongly pinned anharmonic
oscillators subject to a non-equilibrium random forcing.  Assuming
that the stationary state is approximately Gaussian, we first derive a
stationary Boltzmann equation. By localizing the involved resonances,
we next invert the linearized collision operator and compute
the heat conductivity. In particular, we
show that~the Gaussian approximation yields a finite conductivity
$\kappa\sim\frac{1}{\lambda^2T^2}$, for
$\lambda$ the anharmonic coupling strength.
\end{abstract}

\pacs{05.70.Ln,05.60.Cd,05.20.Dd,65.40.-b,66.70.+f}
\maketitle
\noindent
The analytical derivation of the Fourier's law of heat conduction in extended Hamiltonian systems remains to this day a challenging question. See \cite{livi,bonetto} for a review of various aspects of the problem. A qualitative explanation for Fourier law has been provided by Peierls for the case of crystalline solids.
Peierls idea essentially consists in considering the lattice excitations of the crystal, the phonons, as particles interacting weakly through the nonlinear forces acting between the atoms.  By analogy with the case of dilute gases, a Boltzmann equation for phonons has been derived and refined over the years, see \cite{Spohn} and references therein.  This approach yields a qualitative comprehension of the heat transfer in terms of collisions between phonons.  The explicit computation of the heat conductivity amounts next to invert  the collision operator between phonons.  To this day, this computation has not been performed. This is due to the fact that the localization of the resonances in the combination of frequencies of the phonons involved in the collision is a notoriously delicate problem.  In this Letter, we solve this problem in the case of a strong harmonic pinning.

Crystalline solids submitted to a heat flow are often modelized by a
chain of oscillators where each oscillator move around some
equilibrium position located on a regular lattice.  Each oscillator
interact with its nearest neigbours and also possibly with some
substrate.  The chain is coupled to heat baths at different
temperatures by some random forcing and friction acting only on the
particles at the boundaries of the chain.  Although  mathematical
properties of such systems have been extensively studied, see
\cite{Rey} for a review, very little is known analytically about their
physical properties, except in cases where the interactions are
quadratic, \cite{rieder}.
Numerous numerical simulations have been performed,
see \cite{livi} and references therein,
strongly supporting that Fourier law is verified in 
pinned anharmonic chains.  In particular, the conductivity of our model has been studied numerically \cite{Baowen,Aoki} and shown to be finite, although with a different scaling in temperature than what the Gaussian assumption of the Peierls theory predict, as we will see below.

As we want to concentrate on the analysis of the operator of collision between phonons,  we derive a stationary
Peierls-Boltzmann equation for a translation-invariant chain. The
equation for the evolution of the average energy current $J_k$ carried
by the mode of wavenumber $k$ of the chain will have the form,
\qq
\frac{d}{dt}J_k(t)=-\gamma J_k(t)+\lambda {\cal N}_k(t) + S_k.
\label{rough}
\qqq
The first term on the RHS of this equation is a damping term coming
from the coupling of the chain to an external friction represented by
the coefficient $\gamma$.  The second term comes from the Hamiltonian
evolution and vanishes when the anharmonic part of the interaction is
zero, i.e., when $\lambda=0$.  The third term represents an external
"creation" of current at constant rate.  It modelizes the local
temperature gradient in the non-periodic chain.  Equation
(\ref{rough}) will therefore yield a balance equation for the current
carried by the mode $k$ in the stationary state.  We will show that
under the assumption that the stationary state is approximately
Gaussian, ${\cal N}_k$ is analogous to a friction term, yielding in
the end a conductivity $\kappa\sim\frac{1}{\lambda^2T^2}$.  From a
mathematical point of view, there is an important difference between
the part coming from the external friction and the part coming from
${\cal N}_k$.  Whereas the friction acts as a diagonal operator on the
vector $J$, we will see that ${\cal N}_k\sim A(J)_k$ where ${A}$ is a
non-diagonal linear operator.  This is the collision operator of the
linearized Boltzmann equation for phonons. The properties of this
operator are essentially governed by resonances in the four-body
interactions between phonons, which we analyze in the limit of a large
harmonic pinning strength $\mu$.  In particular, we show explicitly
that the origin of the finite conductivity is to be found in
$\mu$-dependent resonances corresponding to {\sl umklapp}
processes. As we concentrate only on the effect of the
term coming from the nonlinearity ${\cal N}_k$, the
source term $S_k$ is put by hand.  However, in the non-translation
invariant case, it comes naturally from the transport term of the
equations and is roughly given by $\omega^{-1}(k)\nabla\omega(k)\nabla
T$ where $\omega(k)$ is the dispersion relation and $\nabla T$ is the
local temperature gradient, see~\cite{Spohn}.

\vskip7pt
\noindent{\bf The Model.}
We consider particles located on a periodic lattice described by their positions and momenta coordinates $(q_i,p_i)$ with $i\in{1,\ldots,N}$. For the sake of notational simplicity, we assume that $N$ is divisible by $4$.  The dynamics of the particles is governed by the Hamilton equations and an external random forcing.  The Hamiltonian that we consider is of the form
\qq
H(\underline q,\underline p)=\sum_{i=1}^N 
\big[\frac{p_i^2}{2}+\omega^2\mu^2\frac{q_i^2}{2}+\frac{\lambda}{4}q_i^4+\frac{\omega^2}{2}(q_i-q_{i-1})^2\big].
\label{Hamilton}
\qqq
The anharmonic interaction is only in the on-site part of the Hamiltonian.  The same analysis as found below may be carried out when the interaction between nearest neighbours contains also a quartic part.  The fact that there is an on-site (large) harmonic interaction makes the distinction between the two cases irrelevant. 
We first consider the harmonic case $\lambda=0$. 
We introduce the  Fourier coordinates for the periodic linear chain by $Q_k=\frac{1}{\sqrt{N}}\sum_{j=1}^{N} e^{i \frac{2\pi}{N}k j}q_j$,with $-N/2+1\leq k\leq N/2$. The $P_k$ coordinates are defined in a similar fashion.
We recall for further purposes that $Q^*_k=Q_{-k}$ and $P^*_k=P_{-k}$.
In the complex coordinates,
\qq
A^{\pm}_{k} = \frac{1}{\sqrt{2\omega_k}}(P_k \pm i \omega_k Q_k),
\label{phon}
\qqq
with $\omega^2_k=\omega^2(\mu^2+4\sin^2(\frac{\pi k}{N}))$,
the Hamilton equations for the linear periodic chain read
\qq
d A^{\pm}_{k}=\pm i \omega_kA^{\pm}_{k} dt.
\label{eqs}
\qqq
Those equations give the temporal evolution of the amplitudes of the waves with wave number $k$ traveling through the chain in the positive or negative direction.
Our (non-equilibrium) model is defined in the linear case by the system of stochastic equations
\qq
d A^{\pm}_k=\pm i \omega_k A^{\pm}_k dt -\frac{\gamma}{2} (A^{+}_k+A^{-}_k) dt + dW^\pm_k\,,
\label{eqp}
\qqq
where the Wiener processes $W^\pm_k$ satisfy the relations
\qq
d(W^{s}_k W^{s'}_{k'})\equiv I^{s,s'}_{k,k'}dt\equiv 2\Bigl[\frac{\gamma T}{\omega_k}+(s-s')\tau\alpha(\frac{\pi k}{N})\Bigr]\delta(k+k')dt.
\label{Ito}
\qqq
In (\ref{eqp}), the term in $\gamma$ is the usual friction term acting
only on the momenta $P_k$.  The Wiener processes represent the random
injection of energy in the system.  
The first term in (\ref{Ito}) is the usual term
giving the equilibrium dynamics.
In the second term of (\ref{Ito}), $\alpha$ is an
odd, dimensionless function of period $\pi$.
Ultimately, the choice of
physically relevant $\alpha$ is fixed by the effect of the local
temperature gradient in the transport term of the non-translation
invariant chain. 
Finally, we interpret $\tau$ in the second term of (\ref{Ito})  as the non-equilibrium
parameter of our model.  When $\tau$ is zero, the periodic chain is in
thermal equilibrium at temperature $T$.  When $\tau\neq 0$, a
stationary heat current proportional to $\tau$ is created in
the chain. This is achieved, via the factor $s-s'$, by exciting more the waves travelling in
one direction than the ones travelling in the opposite direction.

\goodbreak
In order to see those points, we recall first that the energy current in the periodic chain reads
\qq
J=\frac{\omega^2}{N}\sum_{k =-N/2+1}^{N/2} \sin {(2\pi k/N)}\,\Im\left(Q_{-k} P_{k}\right),
\label{current}
\qqq
where $\Im(z)$ denotes the imaginary part of a complex $z$.
We define the two-point correlation functions in the stationary state,
\qq
\Phi^{s,s'}_{k,k'}\equiv \left<A^{s}_{k}A^{s'}_{k'}\right>\,,
\label{phi}
\qqq
where $\left<.\right>$ denotes the average w.r.t.~the stationary measure.
Since $\Im (\left<Q_{-k}P_{k}\right>)={1\over2}(\Phi^{+ -}_{k,-k}-\Phi^{- +}_{k,-k})={1\over2}(\left<|A^+_k|^2\right>-\left<|A^-_k|^2\right>)$, we get
\qq
\left<J\right>=  \frac{\omega^2}{2N}\sum_{k =-N/2+1}^{N/2} \sin {(2\pi k/N)}\left[\left<|A^+_k|^2\right>-\left<|A^-_k|^2\right>\right].
\label{curr}
\qqq
This identity expresses the fact that the energy current is the difference of the current carried by waves traveling in the positive and negative directions.  From (\ref{eqp}), we derive the equation for the two-point correlation functions,
\qq
i(s\omega_k+s'\omega_{k'})\Phi^{s,s'}_{k,k'}=\gamma\Phi^{s,s'}_{k,k'}+\frac{\gamma}{2}(\Phi^{-s,s'}_{k,k'}+\Phi^{s,-s'}_{k,k'})-\hf I^{s,s'}_{k,k'}
\label{sp}
\qqq
with $I^{s,s'}_{k,k'}$ given in (\ref{Ito}).
In the case where $\tau=0$, (\ref{sp}) is readily solved to yield, using (\ref{Ito}),
\qq
\Phi^{s,s'}_{k,k'}=\frac{T}{\omega_k}\,\delta(k+k')\,\delta(s+s').
\label{eq}
\qqq
This is of course equivalent to $\left<|P_k|^2\right>+\omega^2_k\left<|Q_k|^2\right>=2T$ and $\left<P_kQ_l\right>=0$, $\forall k,l$.

When $\tau\neq 0$ and for $l=-k$, one  obtains from (\ref{sp}) and (\ref{Ito}),
\qq
J_k\equiv \Phi^{+ -}_{k,-k}-\Phi^{- +}_{k,-k}=4\tau\gamma^{-1}\alpha(\frac{\pi k}{N}),
\label{cu}
\qqq
which is the balance equation (\ref{rough}) in the stationary state for the harmonic case.
We note that by definition, $J_k=-J_{-k}$.
Using (\ref{curr}) and the fact that $\alpha$ is an odd function, we get $\left<J\right>=2\omega^2\gamma^{-1}\tau\int_{0}^{\pi/2}\alpha(x)\sin(2x)dx$ for large $N$. 
As we interpret $\tau$ as a local non-equilibrium parameter, the conductivity in the harmonic case, $\kappa\equiv \left<J\right>/\tau \sim \gamma^{-1}\omega^2$,  behaves as in the harmonic chain coupled to self-consistent reservoirs \cite{Bon}.  It reflects the fact that when $\lambda=0$ in the balance equation (\ref{rough}), the only term damping the current carried by each mode comes from the external friction.  We will see below that the anharmonic interactions modify that behaviour of the conductivity. 

Let us now consider the anharmonic case $\lambda\neq 0$. The equations of motion (\ref{eqp}) become
\qq
d A^{\pm}_k=\Bigl[\pm i \omega_k A^{\pm}_k -\frac{\gamma}{2} (A^{+}_k+A^{-}_k) -\frac{i\lambda}{N}R_k\Bigr]dt+ dW^\pm_k\,,
\label{neqp}
\qqq
where
\beqnarray
R_k&=&\sum_{k_1,k_2,k_3}\sum_{s_1,s_2,s_3}\delta(k-k_1-k_2-k_3)\,s_1s_2s_3\qquad\nonumber\\&&\qquad\qquad\quad\quad\qquad L_{k k_1k_2k_3}A^{s_1}_{k_1}A^{s_2}_{k_2}A^{s_3}_{k_3}\,,
\eeqnarray
and $L_{k k_1k_2k_3} =(16\,\omega_k\omega_{k_1}\omega_{k_2}\omega_{k_3})^{-1/2}$. Here and below, unless otherwise specified, the sums are over the $k_i$ such that $-N/2+1\leq k_i\leq N/2$ and $s_i=\pm 1$. 

From (\ref{neqp}), the equations for the $n$-point correlation functions in the stationary state,
\qq
(\Phi^{(n)})^{s_1,\ldots,s_n}_{k_1,\ldots,k_n}\equiv \left<A^{s_1}_{k_1}\ldots A^{s_n}_{k_n}\right>,
\label{npoint}
\qqq
read
\qq
i\bar\omega^{(n)}\Phi^{(n)}=\gamma\Gamma^{(n)}(\Phi^{(n)})+i\frac{\lambda}{N}M^{(n)}(\Phi^{(n+2)})-\hf I^{(n)}(\Phi^{(n-2)}),
\label{hier}
\qqq
where $\bar\omega^{(n)}$ is the combination of frequencies,
$$
(\bar\omega^{(n)})^{s_1,\ldots,s_n}_{k_1,\ldots,k_n}= \sum_{i=1}^{n}s_i\omega_{k_i}\,,
$$
and
\beqnarray
(\Gamma^{(n)}(\Phi^{(n)}))^{s_1,\ldots,s_n}_{k_1,\ldots,k_n}=\frac{n}{2}(\Phi^{(n)})^{s_1,\ldots,s_n}_{k_1,\ldots,k_n}\qquad\qquad\quad\nonumber\\+\,\frac{1}{2}\sum_{i=1}^{n}(\Phi^{(n)})^{s_1,\ldots-s_i,\ldots,s_n}_{k_1,\ldots,k_n}.
\eeqnarray
$I^{(n)}$ gathers the effects of the random forcing, the special case $I^{(2)}$ being given in (\ref{Ito}). The explicit expression of $M^{(n)}$ will be given below  in the relevant cases.
For $n=2$ and $k_1=-k_2=k$, 
one gets from (\ref{hier}) the balance equation in the stationary state,
proceeding as to~obtain~(\ref{cu}),
\qq
\gamma J_k +\frac{i \lambda}{N}\Psi_k =4 \tau\alpha(\frac{\pi k}{N})\,,
\label{center}
\qqq
where $J_k$ has been defined in (\ref{cu}) and
\beqnarray
\Psi_k\equiv(M^{(2)}(\Phi^{(4)}))^{+-}_{k,-k}-(M^{(2)}(\Phi^{(4)}))^{-+}_{k,-k}\qquad\ \,\nonumber 
\\=2i\sum_{k_1,k_2,k_3}\sum_{s_1,s_2,s_3}\delta(k+k_1+k_2+k_3)\,s_1s_2s_3\nonumber &&
\\ L_{k k_1k_2k_3}\Im[(\Phi^{(4+)})^{+s_1s_2s_3}_{k,k_1,k_2,k_3}]\,,\label{MM}
\eeqnarray
with
\qq
(\Phi^{(n\pm)})^{s_1,\ldots,s_n}_{k_1,\ldots,k_n}\equiv(\Phi^{(n)})^{s_1,\ldots,s_n}_{k_1,\ldots,k_n}\pm(\Phi^{(n)})^{-s_1,\ldots,-s_n}_{k_1,\ldots,k_n}.
\qqq
The remainder of this paper is devoted to solving equation (\ref{center}) for $J_k$.

\vskip7pt
\noindent{\bf Closure and Linearization.}
In this section, we derive an explicit expression for $\Psi$ in
terms of $J$. The first step is to approximate the $n$-point
correlation functions (\ref{npoint}) by products of $\Phi^{(2)}$, turning
(\ref{center}) into a close but nonlinear equation. 
This Gaussian closure assumption on the non-equilibrium stationary
measure
is widely used in various forms, such as in, e.g., the study of dilute gases.
The second step is to linearize the resulting
equation around the equilibrium solution (\ref{eq}). 
This is done in the spirit of taking a small temperature gradient  or, in our case, a small parameter $\tau$.

Closing as described above the expression (\ref{MM}) for $\Psi$
yields zero after linearization, so that equation (\ref{center})
is identical to equation (\ref{cu}) for the harmonic case. We thus
first express $\Phi^{(4)}$ in (\ref{MM}) in terms of $\Phi^{(6)}$. 
Inverting (\ref{hier})
for $\Phi^{(4)}$, one can show that 
at lowest order in $\gamma$
\qq
\Im[(\Phi^{(4+)})^{+s_1s_2s_3}_{k,k_1,k_2,k_3}]=-\frac{\lambda}{N}(\Omega_\gamma M^{(4)}(\Re[\Phi^{(6-)}]))^{+s_1s_2s_3}_{k,k_1,k_2,k_3}\,.
\label{P4}
\qqq
where
$$
(\Omega_\gamma)^{s_1s_2s_3s_4}_{k_1,k_2,k_3,k_4}=\frac{2\gamma}{4\gamma^2+(s_1 \omega_{k_1}+s_2\omega_{k_2}+s_3\omega_{k_3}+s_4\omega_{k_4})^2}\,,
$$
In (\ref{P4}), we have
dropped a term containing $\Im[\Phi^{(6+)}]$ as well as the contribution from the It\^o term $I^{(4)}$, since they vanish later in the computation.  
We note that when $\gamma\rightarrow 0$, 
\qq
(\Omega_\gamma)^{s_1s_2s_3s_4}_{k_1,k_2,k_3,k_4}\rightarrow \delta(s_1 \omega_{k_1}+s_2\omega_{k_2}+s_3\omega_{k_3}+s_4\omega_{k_4}).
\label{delta}
\qqq
We next plug (\ref{P4}) into (\ref{MM}) and use the Wick formula to express $\Phi^{(6)}$ in terms of $\Phi^{(2)}$.  Linearizing around the equilibrium solution
(\ref{eq}) results in an expression for $\Psi$ which depends solely
on $J$ and contains terms  which may be represented graphically by $12$ topologically distinct diagrams. Those diagrams are structurally similar to the usual diagrams of the equilibrium $\lambda\phi^4$ theory.  The main difference comes from the presence of the quantity which is the unknown of the problem, i.e., the vector $J$.

For large pinning interaction $\mu$, the result of the computation is
(here and below, $c$ denotes combinatorial positive factors),
\qq
\Psi_k=-{c\,T^2\over\omega^6\mu^6}{i\lambda\over N}
\sum_{l,n}(\Omega_\gamma)^{++--}_{k,l,n,k+l+n}
\bigl[J_k+J_l+J_n-J_{k+l+n}\bigr].
\label{Psi2}
\qqq
The RHS is the linear operator ${A}$ mentioned in the introduction,
that is, the linearized collision operator of the Boltzmann equation.
We remark that, in quantum-mechanical language, only the interactions
conserving the number of phonons contribute, as the other interactions
are free from resonances.  We will estimate (\ref{Psi2}) in the limit
of large $N$ and for small $\gamma$, more precisely,
$1\gg\gamma/\omega\gg1/N$. We first observe that the main
contributions to the sum will arise from the resonances in
$\Omega_\gamma$, i.e., for lattice points $(l,n)$ near the zeros of
the function $f_x:[-{\pi\over2},{\pi\over2}]^2\rightarrow{\bf R}$,
\beqnarray
f_x(y,z)&=&\sqrt{\mu^2+4\sin^2(x)}+\sqrt{\mu^2+4\sin^2(y)}\nonumber\\&&
-\ \sqrt{\mu^2+4\sin^2(z)}-\sqrt{\mu^2+4\sin^2(x+y+z)}.\nonumber\\&&
\label{f}
\eeqnarray
A careful analysis reveals that for $\mu^2>0$,
the zeros of $f_x$ form three smooth curves. Two of these curves are
obvious and given by $x+z=0$, and $y+z=0$. These resonances do
not contribute to the sum, however, since the combination of $J$'s in (\ref{Psi2}) vanishes for
$k+n=0$ or $l+n=0$.   This corresponds to the so-called {\sl normal}
processes. The third curve, corresponding to the {\sl umklapp} processes, depends on $\mu$ and is
difficult to localize explicitly.
For large $\mu$, it is given 
by $x+y={\pi\over2}+{\cal O}(1/\mu^2)$.
Performing the sum over $l$ in (\ref{Psi2}) in the above-mentioned parameter
regimes thus yields
at lowest order in $\gamma$ and $1/\mu$
(resonances are at $l=-k+N/2$),
\qq
\Psi_k=-i\lambda{c\,T^2\over\omega^7\mu^5}
\sum_{n}{J^-_k+J^-_n
\over
|\sin({\pi(k+n)\over N})\cos({\pi(k-n)\over N})|}\,,
\label{Psi3}
\qqq
where we have defined
$J^-_k\equiv{1\over2}(J_k-J_{k+N/2})$.
Before proceeding to solving equation (\ref{center}),
we note that any vector $J$ can be decomposed as 
$J=J^++J^-$ where $J^-$ as above and 
$J^+_k\equiv{1\over2}(J_k+J_{k+N/2})$ have
the symmetry
properties $J^{\pm}_{k+N/2}=\pm J^{\pm}_{k}$.
It then follows from (\ref{Psi3}) that
$J^+$ does not contribute to $\Psi$, and that $\Psi=\Psi^-$.

\vskip7pt
\noindent{\bf Solution to the Current Equation.}
Recall that $J$ is odd and periodic of period $N$. We first observe
that $J^+$ does not contribute to the average current
$\langle J\rangle$ given by (\ref{curr}). This follows from
$\sum_k\sin({2\pi k/N})J^+_k=0$.
Since, in addition, $J$ is mapped by (\ref{Psi3}) into a vector $\Psi$ with
$\Psi=\Psi^-$ and since
$J^+$ does not contribute to $\Psi$, we need only to consider in equation (\ref{center})
odd forcings $\alpha$ and odd currents $J$ satisfying
$\alpha({\pi k\over N}+{\pi\over2})=-\alpha({\pi k\over N})$ and
$J=J^-$. 
We denote by ${\cal S}^-$ the subspace of such vectors $J$,
in which equation (\ref{center}) becomes
\qq
\gamma J_k+c\,{\lambda^2T^2\over\omega^7\mu^5}{\cal L}(J)_k
=4\tau\alpha({\pi k\over N})\,,
\label{center2}
\qqq
where $\cal L:{\cal S}^-\rightarrow{\cal S}^-$ is given by
\qq
{\cal L}(J)_k={1\over N}\sum_{k'}{J_k+J_{k'}
\over
|\sin({\pi(k+k')\over N})\cos({\pi(k-k')\over N})|}\,.
\label{Delta}
\qqq
We now proceed to analyze the linear operator ${\cal L}$.
The subspace ${\cal S}^-$ has dimension $N/4$, and a basis for ${\cal S}^-$
is given by  ${\cal J}^n_k=\sin(2\pi(2n+1)\frac{k}{N})$, $n=0,\ldots,N/4-1$.
We let ${\cal A}^n\equiv{\cal L}({\cal J}^n)$.
A direct computation shows that the set of ${\cal A}^n$, $n=0,\dots,N/4-1$,
also forms a basis of ${\cal S}^-$ and that ${\cal L}$ is uniformly invertible (in $N$).
This implies that the first term on the LHS of (\ref{center2}) is negligible  
for $\gamma$ small.  
Furthermore, it follows from $\sum_k\sin(\frac{2\pi k}{N}){\cal J}^n_k=0$ for $n\neq 0$
that only  ${\cal J}^0_k$ contribute to the current (\ref{curr}). Therefore,
the only contribution of the noise $\alpha$ to the current is the component of $\alpha$ along ${\cal A}^0$, say $\alpha^0$, where
\qq
{\cal A}^0_k\equiv{\cal L}({\cal J}^0)_k= 2\, {\rm sign}(k)\Bigl(\frac{1}{4}-\Bigl|{\rm sign}(k)\frac{k}{N}-\frac{1}{4}\Bigr|\Bigr).
\label{image}
\qqq
One thus finally obtains from (\ref{curr}),
\qq
\left<J\right>\sim\frac{\omega^9\mu^5}{\lambda^2 T^2}\,\alpha^0\,\tau\,,
\label{final}
\qqq

\vskip7pt
\noindent{\bf Conclusions.} As mentioned in the introduction, the result of \cite{Spohn} shows
that the local gradient of temperature imposes the choice
$\alpha(k)\sim\omega^{-1}(k)\nabla\omega(k)$, that is,
$\mu^{-2}\sin(2\pi k/N)$ for large $\mu$.
It has a non-zero component
along ${\cal A}^0$ and the corres\-ponding conductivity is thus given by
$\kappa\equiv\left<J\right>/\tau\sim\frac{\omega^9\mu^3}{\lambda^2
T^2}$.
Our explicit treatment of the resonances and the inversion of the
linearized Boltzmann operator allows to justify rigorously the
physical picture that the mechanism responsible for the normal
conductivity in pinned anharmonic chains is the four-body {\it umklapp} collisions
between phonons.  
Indeed, considering a cubic instead of a quartic interaction
in the Hamiltonian would yield an expression analogous to
(\ref{delta}) but with a combination of only three frequencies. 
The latter is always non-zero when $\mu\neq0$, 
leading to an infinite conductivity.
We finally point out that when the pinning goes to zero, the localization of
the resonances and the inversion of the linearized collision operator
are much more difficult and there is no reason for expression
(\ref{final}) to be valid.  
\vskip7pt
\noindent{\bf Acknowledgments.} We are grateful to A.~Kupiainen and
J.~Bricmont for numerous discussions.
A.S.~was supported by the Swiss National Science Foundation.


\begin{thebibliography}{10}

\bibitem{Aoki} K.Aoki, D.Kusnezov :{Non-equilibrium Statistical Mechanics of Classical Lattice $\phi^4$ field theory.} Annals of Physics  {\bf 295}  50 (2002)


\bibitem{Bon}F.~Bonetto, J.~L.~Lebowitz, J.~Lukkarinen: {Fourier's law for a harmonic crystal with self-consistent stochastic reservoirs.}
J.~Stat.~Phys.~{\bf 116}, 783--813 (2004).

\bibitem{bonetto} F.~Bonetto, J.~L.~Lebowitz, L.~Rey-Bellet : {Fourier's law: a challenge to theorists.} {\it Mathematical Physics 2000}, 128--151, Imp.~Coll.~Press, 2000.

\bibitem{Baowen} B. Hu, B. Li, H. Zhao: {Heat conduction in
  one-dimensional nonintegrable systems.}
Phys.~Rev. E {\bf 61}, 3828--3831 (2000).




\bibitem{livi} S.~Lepri, R.~Livi, A.~Politi: 
{Thermal conduction in classical low-dimensional lattices.} 
 Phys.~Rep.~{\bf 377}, 1--80 (2003).

\bibitem{Rey} L.~Rey-Bellet: {Nonequilibrium statistical mechanics of open classical systems.} Proceeedings of ICMP (2003).

\bibitem{rieder} Z.~Rieder, J.~L.~Lebowitz, E.~Lieb:
{Properties of a harmonic crystal in a stationary nonequilibrium state.} J.~Math.~Phys.~{\bf 8}, 1073--1078 (1967).

\bibitem{Spohn} H.~Spohn :{The phonon Boltzmann equation, properties and link to weakly anharmonic lattice dynamics.} arXiv:math-ph/0505025.

\end{thebibliography}
\end{document}